\documentstyle[12pt]{article}
\newcommand{\be}{\begin{equation}} 
\newcommand{\ee}{\end{equation}} 
\newcommand{\bea}{\begin{eqnarray}} 
\newcommand{\eea}{\end{eqnarray}}
\newcommand{\q}{\quad}
\newcommand{\bc}{\begin{center}}
\newcommand{\ec}{\end{center}}
\newcommand{\ol}{\overline}
\newcommand{\p}{\prime}

\newcommand{\Dsp}{\displaystyle}

\begin{document}

\vskip 1in

\bc
{\Large \bf Qualitative Properties of Scalar-Tensor Theories of Gravity} \\[2mm]
\ec

\bc
A.A. Coley\\
Department of Mathematics and Statistics\\
Dalhousie University, Halifax, Nova Scotia  B3H 3J5\\ 
\ec

PACS: 98.80 Cq. 04.50.+h

\vskip .5in

\section*{Abstract} 
 
The qualitative properties of spatially homogeneous
stiff perfect fluid
and minimally coupled massless scalar field models
within general relativity are discussed.  Consequently, by exploiting 
the formal equivalence under conformal transformations and 
field redefinitions of certain classes of theories of gravity, the 
asymptotic properties of spatially homogeneous models in a class of 
scalar-tensor theories of 
gravity that includes the Brans-Dicke theory can be 
determined. For example, exact solutions are presented, which
are analogues of the general relativistic Jacobs stiff perfect
fluid solutions and vacuum plane wave solutions, which act as
past and future attractors in the class of spatially homogeneous 
models in Brans-Dicke theory.

\section{Introduction}
 
Scalar-tensor theories of gravitation, in which gravity is mediated by a long-range
scalar field in addition to the usual tensor fields present in Einstein's 
theory, are the most natural alternatives to general relativity (GR).
Scalar-tensor theories of gravity were originally motivated by the desire to
incorporate a varying Newtonian gravitational constant, $G$, into GR, where a
varying $G$ was itself postulated for a variety of observational and theoretical
reasons (cf. Barrow, 1996).  Indeed, the simplest Brans-Dicke theory of gravity
(BDT; Brans and Dicke, 1961), in which a scalar field, $\phi$, acts as the source
for the gravitational coupling with $G\sim\phi^{-1}$ , was essentially motivated by
apparent discrepancies between observations and the weak-field predictions of
GR.  More general scalar-tensor theories with a non-constant BD parameter,
$\omega(\phi)$, and a non-zero self-interaction scalar potential, $V(\phi)$, have been 
formulated, and the solar system and astrophysical constraints on these
theories, and particularly on BDT, have been widely studied (Will, 1993; see
also Barrow and Parsons, 1997).  Observational limits on the present value of
$\omega_0$ need not constrain the value of $\omega$ at early times in more general scalar-tensor
theories (than BDT).  Hence, more recently there has been greater focus
on the early Universe predictions of scalar-tensor theories of gravity, with 
particular emphasis on cosmological models in which the scalar field acts as a
source for inflation (La and Steinhardt, 1989; Steinhardt and Accetta, 1990).

There are many exact cosmological solutions known in BDT.  The earliest flat
isotropic and homogeneous Friedmann-Robertson-Walker (FRW) exact BDT
solutions presented were the vacuum solutions of O'Hanlon and Tupper (1972)
and the special class of power-law perfect fluid solutions of Nariai (1968)
with  $p = (\gamma-1)\rho$ , where  $\gamma$  is a constant.  The general solutions can be
found for all $\gamma$; exact  zero-curvature solutions were given by
Gurevich et al. (1973) and the curved FRW models were presented by Barrow
(1993) (these solutions are surveyed in Holden and Wands, 1998).

A phase-space analysis of the class of FRW models was performed by Kolitch
and Eardley (1993) and was improved upon by Holden and Wands (1998) who
presented all FRW models in a single phase plane (including those at
"infinite" values via compactification).  It was found that typically at
early times $(t \rightarrow 0)$ the BDT solutions are approximated by vacuum solutions
(i.e., the O'Hanlon-Tupper FRW vacuum solutions) and at late times $(t \rightarrow \infty)$
by matter-dominated solutions, in which the matter is dominated by the BD
scalar field (e.g., the power-law Nariai solutions).  Particular attention
was focussed on whether inflation occurs and whether models have an initial 
singularity.

A variety of exact spatially homogeneous but anisotropic BDT solutions have
also been found (Brans and Dicke, 1961; Nariai, 1972; Belinskii and 
Khalatnikov, 1973; Ruban, 1977; Lorenz-Petzold, 1984 and citations therein,
Mimosa and Wands, 1995b).  Various partial results concerning the 
asymptotic behaviour of Bianchi models in BDT have also been discussed.  For
example, Chauvet and Cervantes-Cota (1995) studied the possible isotropization
of special classes of Bianchi models and Guzman (1997) presented a proof
of the cosmic-no-hair theorem for ever-expanding spatially homogeneous BDT
models with matter and a positive constant vacuum energy-density.  However,
there is no comprehensive and definitive discussion of the qualitative 
properties of anisotropic models in BDT.

Exact perfect fluid solutions in scalar-tensor theories of gravity with a
non-constant BD parameter $\omega(\phi)$ have been obtained by various authors; the
isotropic FRW vacuum and radiation solutions of Barrow (1993), which utilized
the techniques of Lorenz-Petzold (1984), were generalized in the 
zero-curvature case to the more general perfect fluid case with a linear 
barotropic equation of state (satisfying $0 \le \gamma \le 4/3$ but including the 
important case of dust $\gamma = 0$) by Barrow and Mimosa (1994) and to stiff fluids
in addition to vacuum and radiation for curved models by Mimosa and Wands
(1995a).  A variety of inflationary and non-inflationary solutions were
obtained.  This work was extended in a systematic study of the qualitative
analysis of curved FRW models with a specific form for $\omega(\phi)$ by Barrow
and Parsons (1997); in particular, the question of whether a given 
scalar-tensor theory solution can approach GR in the weak-field limit at late times
was addressed.  This work was further generalized by Mimosa and Wands (1995b) 
to various special anisotropic Bianchi models for both BDT and scalar-tensor
theories with a particular form for  $\omega(\phi)$; both exact solutions were obtained
and the asymptotic limits of the solutions, including their possible
isotropization, were studied.  The qualitative properties of both isotropic
and special anisotropic scalar-tensor theory models was also studied by
Serna and Alimi (1996).  Isotropization and inflation in anisotropic 
scalar-tensor theories was
discussed earlier by Pimentel and Stein-Schabes (1989).  In summary, there
exist a multitude of partial results on the possible qualitative behaviour
of cosmological models in scalar-tensor theories, where the details of their
asymptotic properties depend on the particular functional form of $\omega(\phi)$ 
assumed [for more references see Wands and Mimosa (1995b) and other papers
cited above].

Scalar-tensor theories with a ``free'' scalar field are perhaps not well motivated
since, often, quantum corrections produce interactions resulting in a non-trivial
potential $V(\phi)$.  More general scalar-tensor theories including a
non-zero scalar potential, and in particular their qualitative properties,
have also been studied (see Billyard et al, 1998, and references therein).

Scalar-tensor theory gravity is currently of great
interest particularly since such theories occur as the low-energy
limit in supergravity theories from string theory (Green et al., 1988) and other
higher-dimensional gravity theories (Applequist et al., 1987).
Indeed, superstring theory is currently the 
favoured candidate for a unified theory of the fundamental
interactions that include gravity, and as such ought to
describe the evolution of the very early Universe.  
In fact BDT, which is the simplest scalar-tensor theory, originated from
taking seriously the scalar field arising in Kaluza-Klein compactification
of the fifth dimension.
Superstring theory 
leads to a variety of new cosmological possibilities including the so-called 
`pre-big-bang'
scenario (Veneziano, 1991; Gasparini and Veneziano, 1993),
and cosmology is the ideal setting in which to study possible
stringy effects.

Lacking a full non-perturbative formulation which 
allows a description of the early Universe close to the Planck
time, it is necessary to study classical cosmology prior
to the GUT epoch by utilizing the low-energy effective
action induced by string theory.  To lowest order in the 
inverse string tension the tree-level effective action in four-dimensions
for the massless fields includes the non-minimally coupled graviton, 
the scalar dilaton
and an antisymmetric rank-two tensor, hence generalizing 
GR (which is presumably a valid description at late, post-GUT,
epochs) by including other massless fields. Additional
fields, depending on the particular superstring model, are 
negligible in this low-energy limit and can be assumed to be frozen,
and hence the massless bosonic sector of (heterotic) 
string theory reduces generically to a four-dimensional scalar-tensor
theory of gravity.  As a result, BDT includes the dilaton-graviton
sector of the string effective action as a special case
$(\omega =-1)$ (Green et al., 1987). \footnote{Although this result is strictly only 
true in the absence of coupling to other matter fields, it remains valid at 
least for 
the massless fields appearing in the low-energy effective string action.}

A variety of exact string-dilaton cosmological solutions
have been found.  These include spatially
homogeneous models (both Bianchi and Kantowski-Sachs
models and their isotropic specializations)  and more 
recently inhomogeneous models (see Barrow and Kunze, 1998, and Lidsey, 1998, and 
references within).
In addition,  Clancy et al. (1998) have begun an investigation of 
the qualitative properties of a class of anisotropic 
Bianchi models within the context of four-dimensional low-energy
effective bosonic string theory.  Applications to string theory of techniques 
developed to study scalar-tensor gravity have
been discussed in the isotropic case by 
Copeland et al. (1994) and in the anisotropic case by Mimosa
and Wands (1995b).

As noted above these results are only partial results obtained by treating 
various special 
cases.  In this paper we shall extend this work
and present results on the general asymptotic 
properties of spatially homogeneous cosmological models in BDT
(and in more general scalar-tensor theories of gravity).  To our
knowledge the only previous generic results in BDT are the investigation of 
the asymptotic
character of solutions close to the cosmological singularity by Belinskii and 
Khalatnikov (1973), 
the study of mixmaster behaviour by Carretero-Gonzales et al. (1994)
and the cosmic-no-hair theorem results of Guzman (1997).

In the next section we establish the formal equivalence 
between stiff perfect fluid models in  GR and 
cosmological models in a class of scalar-tensor theories of 
gravity (including BDT) under conformal
transformations and field redefinitions.  In section 3 we 
then discuss the known asymptotic properties of stiff perfect
fluid models in GR; summarizing these  results:

\begin{enumerate}

\item   For all models (Bianchi models of classes A and B), a subset of the Jacobs
Disc, which consists of exact self-similar {\it Jacobs} \ stiff fluid solutions 
(corresponding to singular points of the 
governing system of autonomous ordinary differential equations), 
is the {\it past} attractor.

\item  As regards {\it future} evolution, all stiff models behave
like {\it vacuum} models with the following exceptions:

\begin{enumerate}
\item[(i)]  Bianchi I models, all of which are exact Jacobs solutions.

\item[(ii)]  Bianchi  II models, which are future asymptotic to 
another subset of the Jacobs Disc.
\end{enumerate}

For Bianchi models of types VI$_0$ and VII$_0$ the future asymptote 
is a flat Kasner model, as in the case of vacuum models.  The 
Bianchi VIII models do not have a self-similar future asymptote;
these ever-expanding stiff models are the only models for which 
this is the case.
\end{enumerate}
 
 \vskip -.25in
In section 3 we also discuss stiff perfect fluid models in GR 
with an additional non-interacting perfect fluid or a cosmological 
constant.  Massless scalar field models in GR are subsequently discussed.
In section 4 we discuss the qualitative properties 
of spatially homogeneous models in a class of scalar-tensor theories
of gravity. This is done by exploiting the formal 
equivalence of these theories with GR and utilizing the 
results of section 3.  We shall concentrate on BDT.  In particular,
we shall present some exact BDT solutions, including
analogues of the general relativistic Jacobs stiff perfect fluid solutions and 
vacuum solutions (and especially a Bianchi type VII$_h$ plane 
wave solution) alluded to above,
which act as past and future attractors in the class of 
spatially homogeneous BDT models.  The asymptotic 
properties of models in the class of scalar-tensor theories
of gravity under consideration can then
be easily determined.  The qualitative properties of 
more general scalar-tensor theories, including those with
a non-zero scalar potential, can be studied in a similar
way (cf. Billyard et al., 1998).

\setcounter{equation}{0}
\section{Analysis} 

A class of scalar-tensor theories, formally equivalent
under appropriate conformal transformations and 
field redefinitions, are given by the action (in the
{\it Jordan} frame) (cf. Mimosa and Wands, 1995b)
\be
\ol{S} = \int \sqrt{-\ol{g}} \left[ \phi \ol{R} -\frac{\omega 
(\phi)}{\phi} \ol{g}^{ab} \phi_{,a} \phi_{,b} +  2\ol{L}_m \right] d^4  x 
\ee 
where $L_m$ is the Lagrangian for the matter fields, which we 
shall assume corresponds to a comoving (i.e., the velocity of matter, $u^a$, 
is parallel to the unit normal to the spatial hypersurface) perfect 
fluid with energy
density $\ol{\rho}$ and pressure $\ol{p}$.  The main purpose of this 
paper is to study the asymptotic properties of spatially homogeneous 
models in this class of theories.  In particular, we are
interested in the  BDT case in which
$\omega(\phi) = \omega_0$, where $\omega_0$ is a constant.

Under the conformal transformation and field 
redefinition
\bea
g_{ab} & = & \phi \ol{g_{ab}} \q (dt = \pm \sqrt{\phi} d\ol{t})\\
\frac{d \varphi}{d \phi} & = & \frac{\pm\sqrt{\omega(\phi) + 3/2}}{\phi}
\eea
the action becomes (in the {\it Einstein} frame)
\be
S = \int \sqrt{-g} [R - g^{ab} \varphi_{,a} \varphi_{,a} +   2L_m] d^4 x,
\ee  
where 
\be
L_m = \frac{\ol{L}_m}{\phi^2}.
\ee

The action $S$ is equivalent to the action for  GR 
minimally coupled to a massless scalar field $\varphi$
and matter $(L_m)$.  We shall attempt to exploit this
equivalence to   study the asymptotic properties of the 
scalar-tensor theories of gravity with action (2.1).  In 
the spatially homogeneous case under consideration
$\phi = \phi(t)$, and hence under the transformation (2.2) the Bianchi 
type of the underlying model is invariant.  Also, in all
applications here the transformation (2.2) is non-singular
and so the asymptotic behaviour of the scalar-tensor
theories (2.1) can be determined directly from the corresponding
behaviour of the GR models (cf. Billyard et al., 1998)

In the scalar-tensor theory (2.1), the energy-momentum
of the matter fields is separately conserved. In the Einstein
frame this is no longer the case (although the overall energy-momentum
of the combined scalar field and matter field is, of course, conserved).  
Indeed (Mimosa
and Wands, 1995b),
\be
\nabla^a T_{ab} = - \frac{1}{2} \frac{\phi_{,b}}{\phi} T^a_a.
\ee
Defining
\bea
\rho & = & \frac{\ol{\rho}}{\phi^2}, \nonumber\\
{} \\
p & = & \frac{\ol{p}}{\phi^2}, \nonumber 
\eea
then in the spatially homogeneous case we obtain
the conservation equation
\be
\dot{\rho} + 3 (\rho + p)H = -Q \dot{\varphi}
\ee
and the Klein-Gordon equation for $\varphi = \varphi(t)$
\be
\ddot{\varphi} + 3 H \dot{\varphi} = Q,
\ee
where 
\be
Q =  \frac{[\rho - 3p]}{\sqrt{2 (3+2 \omega)}}  .
\ee
When $Q \neq 0$, equations (2.8) and (2.9) indicate energy-transfer
between the matter and scalar field.  $Q =0$ when
$T \equiv T^a_a  = 3p - \rho=0$ (i.e., $3 \ol{p} - \ol{\rho} = 0$).  
We shall assume that
the matter satisfies the equation of state \enskip $\ol{p} = 
(\gamma -1) \ol{\rho}$ 
\enskip (i.e., $p = (\gamma -1) \rho)$
where $\gamma$ is a constant.

Finally, defining (Tabensky and Taub, 1973)
\be
\rho_\varphi = p_\varphi = \frac{1}{2} \dot{\varphi}^2,
\ee
so that (2.9) becomes
\be
\dot{\rho}_\varphi + 3 (\rho_\varphi + p_\varphi) H = Q \dot{\varphi}, 
\ee
we see that the massless scalar field is equivalent   
 to a stiff perfect fluid $(\gamma_\varphi =2$).  Hence the model is
equivalent to an interacting two-fluid model,
one fluid of which is stiff (Mimosa and Wands, 1995b).

The study of interacting two-fluid models is very
complicated.  However, there are three special  cases of interest in 
which there is no interaction between the two fluids.  First,
from equation (1.10), $Q = 0$ if $\rho = 3p$ ($\ol{\rho}= 3 \ol{p})$; this can 
occur either for 
vacuum $(\ol{\rho} = \ol{p} =0$; i.e., no matter present) or in the case
of radiation $(\gamma = 4/3)$.  Second, in the case of stiff matter
$(\gamma =2)$, the total energy density and pressure are given by
\be
p_{tot} = p + p_\varphi = \rho + \rho_\varphi = \rho_{tot},
\ee
and so the two-fluid model is equivalent to 
single stiff fluid model satisfying equations (1.13).
Finally, the case in which the matter field is equivalent
to a cosmological constant is tractable.

Although progress is only possible in these very special cases, these cases 
are nonetheless of 
particular physical importance.  For example, cosmological models with
matter are known to be asymptotic to vacuum models in a variety 
of circumstances (Wainwright and Ellis (WE), 1997), and radiation matter 
fields and a cosmological constant
or a vacuum energy density are known to play an important r\^ole 
in the early Universe.  In addition, stiff homogeneous perfect fluids represent
(additional) homogeneous massless scalar fields and can formally model
geometry effects [e.g., the shear scalar in a Bianchi I model in the conformally
transformed Einstein frame behaves exactly like a stiff fluid (Mimosa and 
Wands, 1995b)].
Indeed, the long wave-length modes of a massless scalar field act like a 
stiff fluid, and a stiff fluid may also describe the evolution of an effectively
massless field [including, in the context of superstring cosmology, the 
antisymmetric tensor field which 
appears in the low energy effective action (Copeland et al., 1994)], and if 
present they would be expected to 
dominate at early times in the Universe (over, for example, any short 
wave-length modes or any other matter fields with $\rho > p$).

In the next section we shall study the vacuum
case by reviewing the asymptotic properties of 
stiff perfect fluid spatially homogeneous models
(WE).  Once this is done, we can
determine the scalar field $\varphi$ by integrating equation (2.12)
whence we can determine the asymptotic properties
of the scalar-tensor models from equations (2.2) and (2.3)
(see section 4).  The properties of models in which the matter
field is a stiff perfect fluid can be deduced from the 
results in the vacuum case.  The case in which the
second fluid is radiation will be dealt with in 
subsection 3.2 and the case of a cosmological
constant will be dealt with in subsection 3.3.

\setcounter{equation}{0} 
\section{ Results in General Relativity}

The qualitative properties of orthogonal spatially homogeneous (OSH) 
perfect fluid
models with an equation of state $p = (\gamma -1)\rho$ within GR
have been studied by Wainwright and 
collaborators (see WE and 
references within).  Indeed, the governing equations 
of these models reduce to a (finite) $n$-dimensional 
polynomial system of autonomous ordinary differential
equations.  Utilizing an orthonormal frame approach
and introducing an expansion-normalized (and hence
dimensionless) set of variables, it was shown that 
one differential equation (for the expansion or the 
Hubble parameter) decouples from the remaining
equations, allowing for the study of a ``reduced'' (i.e., 
$(n-1)$-dimensional) system of ordinary differential equations.
In particular, it was proven that all of the singular
points of the ``reduced'' dynamical system correspond
to exact (time-evolving) solutions admitting a 
homothetic vector (Hsu and Wainwright, 1986).  Therefore,
these (transitively) self-similar cosmological models play an
important r\^ole in describing the asymptotic behaviour
of the spatially homogeneous cosmologies.  The dynamics
of the more general Bianchi models is complicated 
by the fact that there exist lower-dimensional attractors
(that are not simple singular points) in Bianchi types VIII
and IX models (which determine their early time behaviour)
and the phase space of models of types VII$_0$, VIII and IX
are not compact (which affects the determination of their late time 
behaviour).

The value $\gamma =2$, of interest in the study
of stiff perfect fluids, is a bifurcation value for
$\gamma$ in the reduced dynamical system; consequently
models with $\gamma =2$ may have different qualitative properties
to models with $\gamma <2$.  A complete discussion 
of the case $\gamma =2$ has yet to be given, so we
begin with a  review of this case.

\subsection{Stiff perfect fluids in GR}

The finite singular points (and their stability) of the reduced 
dynamical system in the case $\gamma =2$ has been 
investigated by Wainwright and collaborators (WE).  Indeed, {\it all} 
non-tilting spatially homogeneous solutions of the 
Einstein field equations with a perfect fluid with $\gamma =2$ (and 
$\rho > 0$) as 
source which admit a four-dimensional similarity group acting simply 
transitively on spacetime are listed in table 9.2 in WE.  In particular, 
the flat isotropic $\gamma =2$ solution (FL) is given by
\bea
ds^2_{FL} & = & -dt^2     + t^{2/3} (dx^2 + dy^2 + dz^2), \nonumber\\
{}\\
\rho & =& \frac{1}{3} t^{-2}, \nonumber
\eea
which is an attractor in the class of isotropic models 
and has important physical applications, 
and the Jacobs stiff perfect fluid solutions $\cal J$   given by
\bea
ds^2_{\cal J} & = & -dt^2 + t^{2p_1} dx^2 + t^{2p_2} dy^2 + t^{2p_3} dz^2,\nonumber\\
{}\\
\rho & = & \frac{1}{2} (1 - p^2) t^{-2}, \nonumber
\eea
where the two essential  parameters are determined by
$$p_1 + p_2 + p_3 = 1,  \q  p^2 \equiv p^2_1 + p^2_2 + p^2_3 < 1. $$
All $\gamma =2$ Bianchi I solutions are Jacobs self-similar solutions and each 
solution corresponds to a singular point on the `Jacobs Disc'.
These  solutions play an important role in describing 
the qualitative properties of classes of Bianchi models. 
Theorem 9.2 in WE states that all known $\gamma =2$
solutions  
correspond to singular points on the Jacobs Disc or 
the (vacuum) Kasner Ring (see WE p.199 for precise definitions of these sets).

In particular, in Wainwright and Hsu (1989) 
OSH models of type A were studied. Although
their analysis was not conclusive, they showed (Proposition
4.1) that there exists a strictly monotonic function on each
of the $\gamma =2$ Bianchi invariant sets.  The singular points occur
on the Jacobs Disc in the Bianchi I invariant set (corresponding 
to Jacobs stiff fluid solution) or the Kasner Ring.  The 
stability of the singular points on the Jacobs Disc, 
complicated by the existence of two zero eigenvalues
(of the five-dimensional set), was discussed on p.1426 in
Wainwright and Hsu (1989), and the stability of those
on the Kasner Ring was discussed on p.1427 where it
was shown that a subset of these points act as 
sources in the various Bianchi type VIII and IX invariant sets.
In addition, OSH models of type B (not including 
the exceptional Bianchi type VI$_{-1/9}$ case) were considered
in Hewitt and Wainwright (1993), where it was shown
(Proposition 5.3) that all such Bianchi models with $\gamma=2$ 
are asymptotic in the past to a Jacob's Bianchi I model
and asymptotic to the future to a vacuum plane wave state.
All self-similar vacuum solutions, including 
the plane wave solutions and various forms of
flat spacetime, are listed in table 9.1 in WE.  For example,
the one-parameter Bianchi type VII$_{h}$ plane wave solution
is given by
\be
ds^2_{PW} = -dt^2 + t^2 dx^2 + t^{2r} e^{2rx} \{e^\beta [\cos v \ dy + 
\sin v \ dz]^2 + e^{-\beta} [\cos v \ dz - \sin v \ dz]^2  \},
\ee
where $v \equiv b(x + ln t)$ and the constants in (3.3) satisfy
$$b^2 \sinh^2 \beta = r(1-r); \ b^2 = r^2/h, \q 0 < r < 1, $$
where $h > 0$ is the group parameter.  The case $r =1 \enskip (\beta = 0)$ 
gives 
the Bianchi VII$_{h}$ version of the Milne model; the Milne form of 
flat spacetime is given by
\be
ds^2_M = -dt^2 + t^2 [dx^2 + e^{2x} (dy^2 + dz^2)].
\ee

As mentioned earlier, all of the singular points correspond 
to transitively self-similar cosmological solutions.  In particular,
the homothetic vectors corresponding to the metrics (3.1)--(3.4)
are given by  
\bea
X_{FL} & = & t \frac{\partial}{\partial t} + \frac{2}{3} 
\left\{x \frac{\partial}{\partial x} + y \frac{\partial}{\partial y} + 
z \frac{\partial}{\partial z}  \right\}, \nonumber\\
X_{\cal J} & = & t \frac{\partial}{\partial t} + (1 - p_1) x 
\frac{\partial}{\partial x} + (1 - p_2) y\frac{\partial}{\partial y} + 
(1 -p_3 ) z\frac{\partial}{\partial z},\nonumber\\
X_{PW} & = & t \frac{\partial}{\partial t} - \frac{\partial}{\partial x} 
+ y\frac{\partial}{\partial y} + z\frac{\partial}{\partial z}, \nonumber\\
X_M & = & t \frac{\partial}{\partial t}.
\eea 

\subsection{Non-interacting perfect fluid and stiff matter}

The analysis of the qualitative properties of two 
{\it non-interacting} perfect fluid OSH models was presented 
in Coley and Wainwright (1992).  The situation of interest
here corresponds to the case in which the first fluid
is stiff $(\gamma_1 =2)$ and the second is radiation $(\gamma_2 = 4/3)$.
Defining the radiation density and the stiff matter density by 
$\rho_r$ and $\rho_s$, respectively, where
\bea
\rho_r & = & \rho_{\gamma = 4/3} \nonumber\\
\rho_s & = & \rho_{\gamma=2} = \rho_\varphi,
\eea
and introducing the new variable $\chi$, defined by
\be
\chi = \frac{\rho_r - \rho_s}{\rho_r + \rho_s};\q -1 \leq \chi\leq 1, 
\ee
from the (separate) conservation laws (2.8) and (2.12) $(Q =0$ from 
equation (2.10)), we obtain the following time
evolution equation for $\chi$:
\be
\chi^\p \equiv \frac{1}{H} \dot{\chi} = 1 - \chi^2 \geq 0.
\ee

Hence, for ever-expanding models ($H > 0)$, $\chi$ is monotonically
increasing with 
\be
\lim_{t \rightarrow 0} \chi = -1, \q \lim_{t \rightarrow \infty} \chi = +1.
\ee
This means that the corresponding cosmological models
evolve from an initial state in which the stiff
fluid $(\chi =-1)$ dominates to a final state in which the 
radiation fluid is dominant $(\chi =+1)$; i.e., the 
asymptotic behaviour of the two-fluid OSH models is 
described by the asymptotic behaviour of the associated single-fluid models.  
Therefore,
the early time behaviour 
of the GR model with a non-interacting stiff fluid and 
radiation or the equivalent massless scalar field model coupled
to radiation, or the associated BDT model with radiation,
can be deduced from the results of the previous subsection.
The late-time behaviour of these models is dominated 
by the radiation.

\subsection{Cosmological constant}

For initially expanding GR spatially homogeneous models with matter and 
a positive cosmological constant,
including the case of a minimally coupled massless scalar field, 
the late time behaviour is determined by the cosmic
no-hair theorem (Wald, 1983); namely, all Bianchi 
models (except a subclass of type IX) are future-asymptotic to de Sitter 
spacetime (see also Coley and 
Wainwright, 1982).

The late time behaviour of ever-expanding spatially homogeneous models in BDT 
with matter and a positive constant vacuum energy  density\footnote{Note that, 
unlike in GR, a cosmological constant is not identical to the presence of a 
vacuum energy in BDT.} 
[where the term $\phi \ol{R}$ in the action (2.1) becomes $\phi (\ol{R} + 
\lambda)$] 
can be determined directly from the cosmic no-hair
theorem results of Guzman (1997); namely, all such models
are future asymptotic to a flat, isotropic, power-law state 
(extended inflation; La and Steinhardt, 1989).\footnote{A 
similar scenario, referred to as hyperextended
 inflation, occurs in scalar-tensor theories of gravity with 
 $\omega(\phi)$ (Steinhardt and Accetta, 1990).}  Isotropization and inflation
in anisotropic scalar-tensor theories was discussed 
earlier by Pimentel and Stein-Schabes (1989).

\setcounter{equation}{0}
\section{Applications}

To study the qualitative properties of OSH
perfect fluid models (with an equation of state $p = (\gamma -1) \rho$) 
within
scalar-tensor theories (with no potential), and particularly
within BDT, expansional-normalized variables can be introduced
and the resulting system of ordinary differential equations can be 
investigated (Billyard et al., 1998).
In BDT it can be shown that again one differential
equation decouples and the ``reduced'' finite-dimensional system
of ordinary differential equations can be analysed (WE); 
the singular points of the reduced dynamical 
system again correspond to exact self-similar solutions
(Coley and van den Hoogen, 1994).  However, here we shall determine
some of the more important qualitative properties directly by utilizing the results
in the earlier sections and noting that solutions corresponding
to singular points of the governing dynamical system can 
act as future and past attractors.

First, we choose coordinates in which the OSH
metric can be written as
\be
ds^2 = -dt^2 + \gamma_{\alpha \beta}(t, x^\gamma) dx^\alpha dx^\beta; 
\q \alpha = 1,2,3.
\ee
At the finite singular points of the reduced 
GR dynamical system (in expansion-normalized variables) it 
can be shown that (Wainwright and Hsu, 1989) $\theta \propto t^{-1}$, 
where $\theta$ is the expansion of the timelike congruences
orthogonal to the surfaces of homogeneity.  Defining the 
Hubble parameter by $\Dsp H = \theta/3$, it follows that
\be
H = H_0t^{-1}.
\ee
Also, at the singular points we have that
\be
\frac{\rho_\varphi}{H^2} = d^2,
\ee
where $d^2$ is a positive constant which can be determined from the
generalized Friedmann equation.  From the energy conservation
equation we then find that $\Dsp H_0 = 1/3$, so that
\be
H = \frac{1}{3} t^{-1}, \q \rho_\varphi = \frac{d^2} {9}t^{-2}.
\ee
Exact cosmological solutions corresponding to the 
singular points must obey equations (4.4).

In addition, these exact OSH solutions are
transitively self-similar (Wainwright and Hsu, 1989); i.e.,
if $g_{ab}$ represents the spacetime metric corresponding to such a 
solution then 
there exists a HV $X$ satisfying
\be
{\cal L} _X g_{ab} = 2 g_{ab},
\ee
where ${\cal L}$ denotes Lie differentiation along $X$.  The HVs 
corresponding to the exact solutions (3.1)--(3.4) were given
by (3.5); we note that $X$ is of 
the form
\be
X = t \frac{\partial}{\partial t} + X^\alpha(x^\gamma) \frac{\partial}
{\partial x^\alpha}
\ee
in all of these four cases.  Indeed, we can show that in the coordinates 
(4.1) any such HV will always be of this form
as follows.

First, since the energy-momentum tensor is of
the form of a perfect fluid with four-velocity $u^a$, it follows
from the Einstein field equations and equation (4.5) that
$${\cal L}_X u^a = -u^a, $$
whence it follows that (Coley and Tupper, 1989)
\be
{\cal L}_X H = -H.
\ee
Now, writing $\Dsp X = X^0 \frac{\partial}{\partial t} + X^\alpha 
\frac{\partial}{\partial x^\alpha}$ and using (4.4),
equation (4.7) implies that $\Dsp X^0 \frac{\partial}{\partial t}(t^{-1}) 
= -t^{-1}$; i.e., $X^0 = t$.
>From (4.1), the $(0 \alpha)$-components of (4.5) then trivially yield 
$X^\alpha = X^\alpha (x^\gamma)$, 
and we obtain the result (4.6).

Next, the scalar field $\varphi(t)$ in the Einstein frame
is related to $\rho_\varphi$ by equation (2.11), so that from (4.4) 
we obtain
\be
\frac{d \varphi}{d t} = \frac{\sqrt{2}d}{3} t^{-1}.
\ee
The scalar field $\phi(t) = \phi(\ol{t}(t))$ in the Jordan frame
is determined by equation (2.3), whence
\be
\frac{1}{\phi} \frac{d \phi}{d t} = 2W (\phi) t^{-1},
\ee
where $\Dsp W(\phi) \equiv \pm \frac{d}{3} (2\omega (\phi) +3)^{-\frac{1}{2}}$. 
In the BDT, where $\phi$ is 
the BD scalar, we have $\omega (\phi) = \omega_0$ and so
\be
W(\phi) = \ol{\omega} \equiv \pm \frac{d}{3} (2 \omega_0 + 3)^{-\frac{1}{2}},   
\enskip  \mbox{a  constant}.
\ee

Now, the scalar-tensor metric $\ol{g}_{ab}$ in the Jordan frame is related
to the GR metric $g_{ab}$ in the Einstein frame by equation (2.2), 
so that
$${\cal L}_X \ol{g}_{ab} = X(\phi^{-1}) g_{ab} + \phi^{-1} {\cal L}_X g_{ab},  $$
whence from equations (4.5), (4.6) and (4.9) we obtain
\bea
{\cal L}_X \ol{g}_{ab} & = & -t \frac{\dot{\phi}}{\phi} \phi^{-1} g_{ab} + 
2 \phi^{-1} g_{ab} \nonumber\\
& = & 2 [1 - W(\phi)] \ol{g}_{ab}.
\eea
Therefore, for a scalar-tensor theory with $\omega = \omega (\phi)$,
$X$ is a conformal Killing vector for the corresponding exact 
solution in the scalar-tensor theory.

In the particular case of BDT (only),
$1 - W(\phi) = 1 - \ol{\omega}$, a constant, and hence $X$ is in fact a HV.
Consequently, the associated exact solution in  BDT,
which can act as a past or future attractor, is again transitively
{\it self-similar}.  The BD scalar field can be 
obtained from equation (4.9), and is given by
(in the time coordinate $t$)
\be
\phi = \phi_0 t^{2 \ol{\omega}},
\ee
where $\phi_0$ is an integration constant.  From equation (4.9) we note that 
the form of $\phi(t)$
in solutions corresponding to the singular points is well defined
for all $t >0$, and hence the conformal transformation
(2.2) is regular and we can therefore deduce the qualitative
behaviour of the scalar-tensor models (as $t \rightarrow 0^+$ and $t \rightarrow 
\infty$)
directly from their GR counterparts.

Finally, we note that for the degenerate case in GR  in which $H$ is a constant
(e.g., de Sitter spacetime or Minkowski spacetime), equation (4.2) and the ensuing 
analysis does not follow.
We note that this is related to the special 
case above in which the BD constant $\omega_0$ is 
such that $\ol{\omega} =1$, whence the vector field $X$ in (4.8) becomes
a Killing vector and the associated GR spacetime
is (four-dimensionally) homogeneous (see Kramer et al., 1980.)

\subsection{Massless scalar field in GR}

The form of the geometry in the exact solutions corresponding to 
singular points  of the governing dynamical system for
stiff perfect fluids was discussed in section 3.  Consequently,
to determine the asymptotic properties of massless scalar
field models in GR we simply need to determine the 
form for the scalar field $\varphi$ in these models, which is 
obtained by integrating equation (4.8), viz.,
\be
\varphi(t) = \varphi_0 + \frac{\sqrt{2}d}{3} \ln t
\ee
(see also Belinskii and Khalatnikov, 1973).

\subsection{Brans-Dicke Theory}

To determine the asymptotic properties of BDT
spatially homogeneous models we shall exploit their
formal equivalence to GR models and use the preceding 
results.  In order to present some
particular results we shall consider the exact GR 
solutions (3.1)--(3.4).

In the BDT (in the Jordan frame) $\phi(t (\ol{t}))$
is given by equation (4.12) and from equations (2.2)
and (4.1) we find that the associated BD metric is given
by (in the time coordinate $t$)
\be
d \ol{s}^2 \equiv d\ol{s}^2_{BD} = -\phi^{-1}_0 t^{-2 \ol{\omega}} dt^2 + 
\phi^{-1}_0 t^{-2 \ol{\omega}} \gamma_{\alpha \beta} (t, x^\gamma) 
dx^\alpha dx^\beta.  
\ee
Defining a new time coordinate $T$ by  
$(\ol{\omega} \neq 1)$
\be
T = ct^{1- \ol{\omega}}, 
\ee 
where $c \equiv \phi^{-\frac{1}{2}}_0 (1 - \ol{\omega})^{-1}$,
we obtain
\be
d\ol{s}^2 = -dT^2 + C^2 T^{-\frac{2 \ol{\omega}}{1 -\ol{\omega}}} 
\gamma_{\alpha \beta} (t (T), x^\gamma) dx^\alpha dx^\beta,
\ee
where $t(T) = (T/c)^{1/(1-\ol{\omega})}$ and $C^2 \equiv 
\phi^{-1}_0 c^{2 \ol{\omega}/(1-\ol{\omega})}$.  Finally, the 
BD scalar field is given by
\be
\phi(T) = C^{-2} T^{\frac{2 \ol{\omega}}{1 - \ol{\omega}}}.
\ee
>From the form of $\phi(T) \ (\ol{\omega} \neq 1)$ we note that 
the transformations
between BDT and GR are non-singular as $T \rightarrow 0^+$ and 
$T \rightarrow \infty$ and hence we can deduce
the asymptotic properties of the BDT models directly from the 
corresponding models in GR.

1:  For the flat isotropic FL metric (3.1), the BD metric
(4.16) becomes (after a constant rescaling of the spatial
coordinates)
\be
d\ol{s}^2_{FL} =- dT^2 + T^{\frac{2(1 - 3\ol{\omega})}{3(1 - \ol{\omega})}} 
(dX^2 + dY^2 + dZ^2).
\ee
We deduce that the 
exact, flat (non-inflationary) isotropic BD solution 
(4.18) and (4.17) is an attractor in the class 
of isotropic models in BDT (see Holden and Wands, 1998).
This vacuum BD solution (where $d^2 =3$ is determined from the 
generalized Friedmann equation) was first obtained by O'Hanlon and Tupper (1972).    
 We note that for large values of $\omega_0$, 
 $\Dsp 2(1 - 3\ol{\omega})/3(1 - \ol{\omega}) 
 \approx \frac{2}{3} \left(1 \pm \frac{d \sqrt{2}}{3 \sqrt{\omega}_0}  \right) 
 \approx \frac{2}{3}$;
indeed, as $\omega_0 \rightarrow \infty$, $\ol{\omega} \rightarrow 0$ and we 
formally recover the 
GR solution in this limit.

2:  For the Jacobs stiff perfect fluid solutions (3.2), the associated 
BD metric (4.16) becomes (after a constant rescaling of each 
spacelike coordinate)
\be
d\ol{s}^2_{\cal J} = -dT^2 + T^{2q_1} dX^2 + T^{2q_2} dY^2 + T^{2q_3} dZ^2, 
\ee
where 
$$q_i = \frac{p_i - \ol{\omega}}{1- \ol{\omega}}; \q \alpha = 1, 2, 3 $$
(Brans and Dicke, 1961; Belinskii and Khalatnikov, 1973).
The Bianchi type I BD solutions (4.19) and (4.17) 
therefore act as  attractors for a variety of OSH
Bianchi models (see subsection 3.1).  In particular, all
non-exceptional, initially expanding Bianchi  type B  
BD models are asymptotic in the past to this BD solution.  The 
metric (4.19) reduces to metric (4.18) in the isotropic case in which 
all of the $q_i$ are equal (i.e., $p_i = 1/3$, $i = 1, 2, 3$).

3:  From subsection 3.1 we can conclude
that all Bianchi models of type B in BDT are 
asymptotic to the future to a vacuum plane wave state.
For example, from (3.3) we obtain the following 
BD Bianchi type VII$_{h}$ plane wave solution (after a constant
rescaling of the `$y$' and `$z$' coordinates)
\be
d\ol{s}^2_{PW} =- dT^2 + D^2 (T^2 dx^2 + T^{2 \frac{(r - \ol{\omega})}
{1 - \ol{\omega}}}  e^{2rx} \{e^\beta [\cos v \ d Y - \sin v \ dZ]^2 + e^{-\beta} 
[\cos v \ dZ - \sin v \ dY]^2  \}),
\ee  
where $D \equiv Cc^{-1/(1 - \ol{\omega})}$ and 
$v = b(x + \frac{1}{1 - \ol{\omega}} ln [T/c])$ (and all other constants
are defined as before).  Metric (4.20) can be simplified 
by a redefinition of the `$x$' coordinate.  The BD
scalar field is given by (4.17).  This exact BD  plane wave solution is believed
to be new.

>From the discussion of the asymptotic properties of 
stiff perfect fluid models in GR given in subsection 3.1 and as 
summarized in the Introduction, we can now deduce the 
asymptotic properties of spatially homogeneous cosmological
models in BDT (in particular, see cases 2 and 3 above).
Indeed, all of the GR results reviewed in the Introduction have BDT analogues.
For example, all orthogonal Bianchi type B BDT models,
except for a set of measure-zero, are asymptotic to the 
future to a vacuum plane-wave state (see, for example,
equation (4.19)).  One immediate consequence of this result,
since Bianchi models of type B constitute a set of positive
measure in the set of spatially homogeneous initial
data, is that a recent conjecture that the initial
state of the pre-big-bang scenario within string theory
generically corresponds to the Milne (flat spacetime)
universe (Veneziano, 1991; Gasparini and Veneziano, 1993; 
Clancy et al., 1998) is unlikely to be valid (recall that
past asymptotic behaviour in pre-big-bang cosmology
corresponds to future asymptotic states in classical
cosmological solutions).

4: Finally, in the special case of de Sitter spacetime
\be
ds^2_{dS} = -dt^2 + e^{2H_0 t} (dx^2 + dy^2 + dz^2),
\ee
we have that $H(t) = H_{0}$, a constant, and we cannot use
the analysis of subsection 4.1 (compare with equation (4.2); 
e.g., equations (4.8) and (4.9) are not valid).  In this case 
we have that
\be
\frac{1}{2} \dot{\varphi}^2 = \rho_\varphi = d^2H^2_0,
\ee
and hence from equations (2.3) and (4.10) we obtain
$$\frac{\dot{\phi}}{\phi} = -6|\ol{\omega}| H_0, $$
whence we find that
\be
\phi(t) = \ol{\phi}_0 \exp  \{-6|\ol{\omega}| H_0 t\},
\ee
where $\ol{\phi}_0$ is an integration constant.
Using (2.2) to obtain the metric in the Jordan frame,
and introducing the new time coordinate $T \propto\exp (3|\ol{\omega}| H_0 t)$,
we obtain (after a constant rescaling of the spatial
coordinates)
\be
d\ol{s}^2_{dS} = -dT^2 +   T^{2 \left(1 + \frac{1}{3|\ol{\omega}|}  \right)}  
(dX^2 + dY^2 + dZ^2), 
\ee
where $\phi(T) \propto T^{-2}$.  This flat, isotropic, power-law
BD solution is clearly inflationary $(1 + 1/3|\ol{\omega}| > 1)$ (cf.
extended inflation, La and Steinhardt, 1989).

\subsection{Scalar-tensor theories}

In the same way we can study the asymptotic properties of scalar-tensor theory models 
with action conformally related to (2.1), where for general $\omega=\omega (\phi), 
\phi(t)$ would be determined from (4.9) (and not given by equation (4.12)).  The 
asymptotic properties of more general scalar-tensor theories can be studied in 
a similar way (cf. Billyard et al., 1998).

\section{Discussion}

The results presented in this paper are the first
concerning the generic asymptotic properties of spatially
homogeneous models in scalar-tensor theories of gravity.  However,
some care is needed in interpreting these results and they 
must be applied in concert with special exact
solutions and the analysis of specific but tractable classes
of models (cf. Mimosa and Wands, 1995b) to build up a 
complete cosmological picture.

First, the GR analysis (in the Einstein frame) is incomplete
in that the phase-space in some Bianchi classes is not compact and 
the Hubble parameter can become zero (and hence 
the expansion-normalized variables become ill-defined). Second,
in scalar-tensor  theories of gravity  (in the Jordan frame)
it is known that there exist solutions which do not have
an initial singularity but have a `bounce' (at which $H =0$) -- typically 
this occurs for negative values of $\omega$; e.g., the
initial singularity is avoided in BDT if $\omega_0 < - 4/3$ (Nariai, 1972).
Since there is always an initial singularity 
in the Einstein frame, such an `avoidance of a singularity' 
is due to the properties of the transformations (2.2) and (2.3);
for example, Mimosa and 
Wands (1995b) describe a set of models that reach an anisotropic singularity in
a finite time in the Einstein frame which correspond to  non-singular 
and shear-free evolution in infinite proper time in the Jordon 
frame.

Consequently, although the asymptotic results presented 
here are generally valid, the full dynamical properties of the 
scalar-tensor models, including their global features and their
physical interpretation, are determined from these asymptotic 
results and the properties of the transformations
and how solutions are matched together to construct the complete
dynamical picture.

ACKNOWLEDGEMENTS

I would like to thank John Wainwright for providing a review of the results concerning the asymptotic
properties of spatially homogeneous stiff perfect fluid models
within general relativity and for comments on the manuscript.
This work was supported, in part, by the Natural Sciences and Engineering Research Council of Canada.

\bc
{\bf References}
\ec

\begin{enumerate}

\item[]  T. Applequist, A. Chodos and P.G.O. Freund, 1987, {\it Modern
 Kaluza-Klein Theories} (Redwood City: Addison-Wesley).
 
 \item[]  J.D. Barrow, 1993, Phys. Rev. D. {\bf 47}, 5329.
 
 \item[]  J.D. Barrow, 1996, MNRAS {\bf 282}, 1397.
 
 \item[] J.D. Barrow and K.E. Kunze, 1998, gr-qc/9807040.
 
 \item[] J.D. Barrow and J.P. Mimosa, 1994, Phys. Rev. D. {\bf 50}, 3746.
 
 \item[]  J.D. Barrow and P. Parsons, 1997, Phys. Rev. D. {\bf 55}, 1906.
 
 \item[]  V.A. Belinskii and I.M. Khalatnikov, 1973, Sov. Phys. JETP 
 {\bf 36}, 591.

\item[] A.P. Billyard, A.A. Coley and J. Ib\'a\~nez, 1998, Phys. Rev. D. {\bf 59}, 023507.

\item[]  C. Brans and R.H. Dicke, 1961, Phys. Rev. {\bf 124}, 925.

\item[]  R. Carretero-Gonzalez, H.N. Nunez-Yepez and A.L. Salas Brito, 1994, 
Phys. Letts. A. {\bf 188}, 48.

\item[]  P. Chauvet and J.L. Cervantes-Cota, 1995, Phys. Rev. D. {\bf 52}, 3416.

\item[]  D. Clancy, J.E. Lidsey and R. Tavakol, 1998, gr-qc/9802052

\item[] A.A. Coley and R.J. van den Hoogen, 1994, in {\it Deterministic Chaos in 
General Relativity}, ed. D. Hobill et al. (Plenum).

\item[] A.A. Coley and B.O.J. Tupper, 1989, J. Math Phys. {\bf 30}, 2618.

\item[] A.A. Coley and J. Wainwright, 1992, Class. Quantum Grav. {\bf 9}, 651.

\item[] E.J. Copeland, A. Lahiri and D. Wands, 1994, Phys. Rev. D. {\bf 50}, 4868.

\item[]  M. Gasperini and G. Veneziano, 1993, Astropart. Phys. {\bf 1}, 317.

\item[]  M.B.  Green, J.H. Schwarz and E. Witten, 1988, {\it Superstring Theory} 
(Cambridge: Cambridge University Press).

\item[]  L.E. Gurevich, A.M. Finkelstein and V.A. Ruban, 1973, Ap. Sp. Sci. {\bf 22},
231.

\item[] E. Guzman, 1997, Phys. Letts. B. {\bf 391}, 267.

\item[] C. Hewitt and J. Wainwright, 1993, Class. Quantum Grav. {\bf 10}, 99.

\item[]  D.J. Holden and D. Wands, 1998, gr-qc/9803021.

\item[]  S.J. Kolitch and D.M. Eardley, 1995, Ann. Phys. (N.Y.) {\bf 241}, 128.

\item[] D. Kramer, H. Stephani, E. Herit and M.A.H. MacCallum, 1980, 
{\it Exact Solutions 
of Einstein's Field Equations} (Cambridge: Cambridge University Press).

\item[] D. La and P.J. Steinhardt, 1989, Phys. Rev. Letts. {\bf 62}, 376.

\item[]  J.E. Lidsey, 1998, preprint.

\item[]  D. Lorenz-Petzold, 1984, in {\it Solutions to Einstein's Equations:  Techniques
and Results}, Proc. of the International Seminar, Retzbach (Germany), 
eds. C.H. Hoenselaers and W. Dietz, Lecture Notes in Physics
volume 205 (Springer-Verlag, Berlin).

\item[] J.P. Mimoso and D. Wands, 1995a, Phys. Rev. D. {\bf 52}, 5612.

\item[] J.P. Mimoso and D. Wands, 1995b, Phys. Rev. D. {\bf 51}, 477.

\item[] H. Nariai, 1968, Prog. Theoret.  Phys. {\bf 40}, 49.

\item[]  H. Nariai, 1972, Prog. Theoret. Phys. {\bf 47}, 1824.

\item[] J. O'Hanlon and B.O.J. Tupper, 1972, Il Nuovo Cimento B {\bf 7}, 305.

\item[] L.O. Pimentel and J. Stein-Schabes, 1989, Phys. Letts. B. {\bf 216}, 27.

\item[]  V.A. Ruban, 1977, Sov. Phys. JETP {\bf 45}, 629.

\item[]  A.A. Serna and J.M. Alimi, 1996, Phys. Rev. D. {\bf 53}, 3074 and 3087.

\item[] P.J. Steinhardt and F.S. Accetta, 1990, Phys. Rev. Letts. {\bf 64}, 2740.

\item[] R. Tabensky and A. H. Taub, 1973, Commun. Math. Phys. {\bf 29}, 61.

\item[]  G. Veneziano, 1991, Phys. Lett. {\bf B265}, 287.

\item[] R. M. Wald, 1983, Phys. Rev. D. {\bf 28}, 2118.

\item[] J. Wainwright and G.F.R. Ellis, 1997, {\it Dynamical Systems in Cosmology}
(Cambridge: Cambridge University Press).

\item[] J. Wainwright and L. Hsu, 1989, Class. Quantum Grav. {\bf 6}, 1409.

\item[]  C.M. Will, 1993, {\it Theory and Experiment in Gravitational
Physics} (Cambridge University Press, Cambridge).
\end{enumerate}

\end{document}